\documentclass[prl,reprint, amsmath,amssymb, aps,preprintnumbers]{revtex4-1}

\usepackage[colorlinks=true,linkcolor=black, citecolor=black,
urlcolor=black]{hyperref}

\usepackage{multirow,graphics}

\usepackage{amstext}
\usepackage{amssymb}
\usepackage{amsmath}
\usepackage{graphicx}
\usepackage{color}

\usepackage{float}
\restylefloat{table}
\restylefloat{figure}

\begin{document}

\newcommand{\IM}{{\rm Im}\,}
\newcommand{\card}{\#}
\newcommand{\la}[1]{\label{#1}}
\newcommand{\eq}[1]{(\ref{#1})}
\newcommand{\figref}[1]{Fig \ref{#1}}
\newcommand{\abs}[1]{\left|#1\right|}
\newcommand{\comD}[1]{{\color{red}#1\color{black}}}

\makeatletter
     \@ifundefined{usebibtex}{\newcommand{\ifbibtexelse}[2]{#2}} {\newcommand{\ifbibtexelse}[2]{#1}}
\makeatother

\newcommand{\footnotea}[1]{\ifbibtexelse{\footnote{#1}}{
\newcommand{\textfootnotea}{#1}
\cite{thefootnotea}}}
\newcommand{\footnoteb}[1]{\ifbibtexelse{\footnote{#1}}{
\newcommand{\textfootnoteb}{#1}
\cite{thefootnoteb}}}
\newcommand{\footnotebis}{\ifbibtexelse{\footnotemark[\value{footnote}]}{
\cite{thefootnoteb}}}

\def\bnu{{\bar{\nu}}}
\def\umu{{\underline{\mu}}}
\def\e{\epsilon}
     \def\bT{{\bf T}}
    \def\bQ{{\bf Q}}
    \def\wT{{\mathbb{T}}}
    \def\wQ{{\mathbb{Q}}}
    \def\ttQ{{\bar Q}}
    \def\tQ{{\tilde \bP}}
        \def\bP{{\bf P}}
    \def\CF{{\cal F}}
    \def\cC{\CF}
     \def\Tr{\text{Tr}}
     \def\l{\lambda}
\def\hbZ{{\widehat{ Z}}}
\def\bZ{{\resizebox{0.28cm}{0.33cm}{$\hspace{0.03cm}\check {\hspace{-0.03cm}\resizebox{0.14cm}{0.18cm}{$Z$}}$}}}
\newcommand{\rb}{\right)}
\newcommand{\lb}{\left(}
\newcommand{\nn}{\nonumber}

\newcommand{\gT}{T}\newcommand{\gQ}{Q}

\newcommand{\nc}{\newcommand}
\nc{\beq}{\begin{equation}}
\nc{\eeq}{\end{equation}}
\nc{\bea}{\begin{eqnarray}}
\nc{\eea}{\end{eqnarray}}
\def\IZ{\mathbb{Z}}
\def\ov{\overline}

\newcommand{\cmark}{\ding{51}}
\newcommand{\xmark}{\ding{55}}

\def\CN {{\cal N}}
\def\om{{\omega}}
\def\sig{{\sigma}}
\def\d{{\delta}}
\def\be{\begin{equation}}
\def\ee{\end{equation}}
\def\bea{\begin{eqnarray}}
\def\eea{\end{eqnarray}}
\def\bes{\begin{subequations}}
	\def\ees{\end{subequations}}
\def\raw{\rightarrow}
\def\Raw{\Rightarrow}
\def\eps{{\epsilon}}
\def\oh{\frac{1}{2}}
\def\CG {{\cal G}}

\newcommand{\bmat}{\left(\begin{array}}
	\newcommand{\emat}{\end{array}\right)}
\newcommand{\uno}{\mathbbm{1}}
\newcommand{\pc}{\mathbbm{C}}
\newcommand{\pr}{\mathbbm{R}}
\newcommand{\bCentering}{\centering}
\newcommand{\bCaption}{\caption}
\newcommand{\sgn}{{\rm sgn}}
\newcommand{\unity}{{\footnotesize\mbox{1\!\!I}}}
\newcommand{\bC}{\mathbb{C}}
\newcommand{\bcross}{\textcolor{blue}{\times}}
\newcommand{\rcircle}{\textcolor{red}{\bigcirc}}
\def\muc{\multicolumn}
\def\bN{\mathbb{N}}
\def\bZ{\mathbb{Z}}
\def\Z{\mathbb{Z}}
\def\R{\mathbb{R}}
\def\C{\mathbb{C}}
\def\Q{\mathbb{Q}}
\def\P{\mathbb{P}}
\def\CK {{\cal K}}
\def\a {\alpha}
\def\b {\beta}
\def\unity{1\!\!{\rm I}}
\def\ov{\overline}
\def\N{\mathbf{N}}
\def\Sym{\mathbf{Sym}}
\def\Anti{\mathbf{Anti}}
\def\Adj{\mathbf{Adj}}
\def\bCc{c.c.}
\def\Tr{\text{Tr}}
\def\IM{\text{Im}\,}
\def\RE{\text{Re}\,}
\def\ov{\overline}
\def\1{{\bf 1}}
\def\2{{\bf 2}}
\def\3{{\bf 3}}
\def\4{{\bf 4}}
\def\6{{\bf 6}}
\def\OR{f\mathcal{R}}
\def\half{\frac{1}{2}}
\def\pp{\uparrow\uparrow}
\def\ap{\uparrow\downarrow}

\def\bnu{{\bar{\nu}}}
\def\umu{{\underline{\mu}}}
\def\e{\epsilon}
     \def\bT{{\bf T}}
    \def\bQ{{\bf Q}}
    \def\wT{{\mathbb{T}}}
    \def\wQ{{\mathbb{Q}}}
    \def\ttQ{{\bar Q}}
    \def\tQ{{\tilde \bP}}
        \def\bP{{\bf P}}
    \def\CF{{\cal F}}
    \def\cC{\CF}
     \def\Tr{\text{Tr}}
     \def\l{\lambda}

\def\IZ{\mathbb{Z}}
\def\ov{\overline}

\newcommand{\fa}{\hat}
\newcommand{\fb}{\MakeUppercase}
\newcommand{\fc}{\tilde }
\newcommand{\Lie}{{\cal L}}

\def\G{\Gamma}
\def\bA{\bar{A}}
\def\bF{\bar{F}}
\def\bP{\bar{P}}
\def\bW{\bar{W}}
\def\bR{\bar{R}}
\def\bS{\bar{S}}
\def\bC{\bar{C}}
\def\bL{\bar{L}}
\def\bpsi{\bar{\psi}}
\def\bl{\bar{\Lambda}}
\def\l{\Lambda}
\def\note#1{ \textcolor{blue}{[#1]}}
\def\rmt{\rm t}
\def\rmz{\rm z}

\def\CN {{\cal N}}
\def\om{{\omega}}
\def\sig{{\sigma}}
\def\d{{\delta}}
\def\be{\begin{equation}}
\def\ee{\end{equation}}
\def\bea{\begin{eqnarray}}
\def\eea{\end{eqnarray}}
\def\bes{\begin{subequations}}
	\def\ees{\end{subequations}}
\def\raw{\rightarrow}
\def\Raw{\Rightarrow}
\def\eps{{\epsilon}}
\def\oh{\frac{1}{2}}
\def\CG {{\cal G}}

\def\hK {{\hat{K}}}
\def\mk {{\mathcal K}}
\def\mg {{\mathcal G}}
\def\mc {{\mathcal C}}
\def\r {{\rho}}
\def\hr {{\hat{\rho}}}
\def\tr {{\tilde{\rho}}}
\def\p {{\partial}}
\def\e {{\epsilon}}
\def\g {{\gamma}}
\def\hg {{\hat{\gamma}}}
\def\te {{\tilde{\epsilon}}}
\def\he {{\hat{\epsilon}}}
\def\tg {{\tilde{\gamma}}}
\def\ts {{\tilde{\sigma}}}
\def\hs {{\hat{\sigma}}}
\def\ts {{\tilde{\sigma}}}
\def\s {{\sigma}}
\def\br {{\bar{\rho}}}
\def\vee {{\varepsilon}}
\def\th {{\theta}}
\def\hth {{\hat{\theta}}}
\def\ph   {{\phi}}
\def\l   {{\langle}}
\def\rr   {{\rangle}}

\newcommand{\cF}{\mathcal{F}}
\newcommand{\cG}{\mathcal{G}}
\newcommand{\cK}{\mathcal{K}}
\newcommand{\cM}{\mathcal{M}}
\newcommand{\cN}{\mathcal{N}}
\newcommand{\cO}{\mathcal{O}}
\newcommand{\cA}{\mathcal{A}}
\newcommand{\cB}{\mathcal{B}}

\newcommand{\ud}{\mathrm{d}}
\newcommand{\ue}{\mathrm{e}}
\newcommand{\ui}{\mathrm{i}}
\newcommand{\us}{\mathrm{s}}
\newcommand{\jq}[1]{{\bf[{\color{red}JQ:} #1]}}

\newcommand{\uD}{\mathrm{D}}
\newcommand{\uL}{\mathrm{L}}
\newcommand{\uR}{\mathrm{R}}

\newcommand{\vp}{\varphi}

\newcommand{\II}{\mathbb{I}}

\newcommand{\vol}{\mathrm{vol}}
\newcommand{\Str}{\mathrm{Str}}

\newcommand{\ul}[1]{\underline{#1}}
\newcommand{\bs}[1]{\boldsymbol{#1}}


\newcommand{\dvd}[1]{{\bf[{\color{red}DP:} #1]}}
\newcommand{\red}[1]{\textcolor{red}{#1}}
\newcommand{\blue}[1]{\textcolor{blue}{#1}}
\newcommand{\black}[1]{\textcolor{black}{#1}}


\title{Global Embedding of Fibre Inflation in Perturbative LVS}


\author{Swagata Bera$^\dagger{^\ast}$, Dibya Chakraborty$^\ddagger$, George K. Leontaris$^\Diamond$, Pramod Shukla$^\dagger$ \footnote{Email: iswagata78@gmail.com, dibyac@physics.iitm.ac.in, leonta@uoi.gr, pshukla@jcbose.ac.in}}


\affiliation{
{$^\dagger$Department of Physical Sciences, Bose Institute,\\
Unified Academic Campus, EN 80, Sector V, Bidhannagar, Kolkata 700091, India}\\
\vspace{0.03cm}
{$^\ast$Department of Integrated Science Education and Research Center, Visva Bharati University, Santiniketan 731235, India}\\
\vspace{0.03cm}
{$^\ddagger$Centre for Strings, Gravitation and Cosmology, Department of Physics, Indian Institute of Technology Madras, Chennai 600036, India}\\
\vspace{0.03cm}
{$^\Diamond$Physics Department, University of Ioannina, University Campus, Ioannina 45110, Greece}
}


\begin{abstract}
We present a global embedding of the Fibre inflation model in perturbative large volume scenario (LVS) in which the inflationary potential is induced via a combination of string-loop corrections and higher derivative F$^4$-effects. The minimal Fibre inflation model developed in the standard LVS generically faces some strong bounds on the inflaton field range arising from the K\"ahler cone constraints, which subsequently creates a challenge for  realising  the cosmological observables while consistently respecting the underlying supergravity approximations. The main attractive feature of this proposal is to naturally alleviate this issue by embedding Fibre inflation in perturbative LVS.
\end{abstract}

\maketitle



\section{Introduction and Motivation}

\vskip-0.4cm
\noindent
In the context of type IIB orientifold compactifications, the Large Volume Scenario (LVS) \cite{Balasubramanian:2005zx} is one of the most popular schemes of moduli stabilisation. The minimal LVS fixes the overall volume (${\cal V}$) of the underlying Calabi Yau threefold (CY$_3$) along with another modulus $\tau_s$ controlling the volume of an exceptional del-Pezzo divisor of the CY$_3$, which is needed to induced the non-perturbative superpotential contributions. Therefore a generic LVS model with 3 or more K\"ahler moduli, arising from CY compactifications with $h^{1,1}({\rm CY})\geq 3$, can generically have flat directions at leading order, which can be promising inflaton candidates with a potential generated at sub-leading order. Based on the geometric nature of the inflaton field and the source of the potential, there are several popular inflationary models in LVS, e.g. Blow-up inflation \cite{Conlon:2005ki,Blanco-Pillado:2009dmu,Cicoli:2017shd}, Fibre inflation \cite{Cicoli:2008gp,Cicoli:2016chb,Cicoli:2016xae,Cicoli:2017axo}, poly-instanton inflation \cite{Cicoli:2011ct,Blumenhagen:2012ue,Gao:2013hn,Gao:2014fva}, and loop-blowup inflation \cite{Bansal:2024uzr}.

Fibre inflation \cite{Cicoli:2008gp}, being a large field model, is among the most attractive class of models realised in LVS. In explicit global embedding attempts for Fibre inflation, it has been observed that the inflaton range is highly constrained due to the K\"ahler cone conditions (KCC) \cite{Cicoli:2017axo}, which subsequently obstruct the inflationary plateau forbidding the generation of sufficient number of e-foldings. A detailed analysis of all the CY$_3$ with $h^{1,1} = 3$ shows that the presence of diagonal del-Pezzo divisor results in a bound on the LVS flat direction \cite{Cicoli:2018tcq}. In fact the reduced moduli space after LVS minimisation turns out to be compact in such cases, and the flat direction $\tau_f$, which is a candidate for inflaton, has the following bound \cite{Cicoli:2018tcq}:
\bea
\label{eq:KC0}
& & c_1 \langle \tau_s \rangle < \tau_f < c_2 f(\langle{\cal V}\rangle, \langle\tau_s\rangle), 
\eea
where $f(\langle{\cal V}\rangle, \langle\tau_s\rangle)\simeq \langle{\cal V}\rangle^{2/3}$ or $\langle{\cal V}\rangle/\langle\tau_s\rangle$, and $c_i\simeq{\cal O}(1)$ and positive. One can quantify the {\it field-range problem} as the fact that typically Fibre inflation needs $\Delta \varphi \simeq {\cal O}(6-8) M_p$ along with $\langle {\cal V} \rangle\simeq{\cal O}(10^3-10^4)$ where $\Delta\varphi$ is the shift in the canonical field $\varphi$ where $\tau_f = e^{2\varphi/\sqrt{3}}$. One may argue to increase the field range by considering larger value for $\langle{\cal V}\rangle$ in order to generate the required e-foldings, however it turns out that subsequently the scalar perturbation amplitude gets accordingly diluted. Moreover, one needs to respect the following mass hierarchy to trust the supergravity approximations
\bea
\label{eq:mass-hierarchy}
& & m_{\rm inf} < H < m_{3/2} < M_{\rm KK} < M_s < M_p,
\eea
where $m_{\rm inf}$ denotes the inflaton mass, $H$ is the Hubble parameter, $m_{\rm 3/2}$ is the gravitino mass, $M_{\rm KK}$ is the lightest KK-mode, and $M_s$ is the string mass. Maintaining the mass-hierarchies on top of field range bound creates a challenge for constructing a trustworthy viable model.

In this work we propose an alternative way of realising Fibre inflation using perturbative LVS framework \cite{Antoniadis:2018hqy,Antoniadis:2018ngr,Antoniadis:2019doc,Antoniadis:2019rkh,Antoniadis:2020ryh,Antoniadis:2020stf, Leontaris:2022rzj, Bera:2024zsk} that shares most of the interesting properties of the standard LVS without the need of non-perturbative corrections. Given that such effects are not necessary in perturbative LVS, the need of exceptional 4-cycles in the CY$_3$ is absent and subsequently the obstruction from the KCC is also naturally avoided.


\section{LVS schemes of moduli stabilisation}

\vskip-0.4cm
\noindent
The F-term scalar potential governing the low energy dynamics of the four-dimensional ${\cal N} = 1$ effective supergravity arising from the type IIB superstring compactifications is encoded in $V = e^{{\cal K}}({\cal K}^{{\cal A} \ov {\cal B}} (D_{\cal A} W) (D_{\ov {\cal B}} \ov{W}) -3 |W|^2)$, where the full K\"ahler potential (${\cal K}$) and the holomorphic superpotential ($W$)  depend on the various moduli, e.g. the complex-structure (CS) moduli $U^i$,  the axio-dilaton $S$ and the K\"ahler moduli $T_\alpha$. For a decoupled structure of ${\cal K}$ such that ${\cal K} \equiv K_{\rm cs} (U^i)+ K(S, T_\alpha)$, the inverse  metric is block-diagonal and $V \equiv V_{\rm cs} + V_{\rm k}$ such that $V_{\rm cs} =  e^{{K}} K_{\rm cs}^{i \ov {j}} (D_i W) (D_{\ov {j}} \ov{W})$ and $V_{\rm k} = e^{{K}}(K^{{A} \ov {B}}(D_{A} W)(D_{\ov {B}} \ov{W}) -3 |W|^2)$ where $A, B \in \{S, T_\alpha\}$. Moduli stabilisation follows a two-step methodology. First, one fixes the CS moduli $U^i$ and the axio-dilaton $S$ by the leading order flux superpotential $W_{\rm flux}$ induced by the 3-form fluxes $(F_3, H_3)$ \cite{Gukov:1999ya} via the  supersymmetric flatness conditions $D_{U^i} W_{\rm flux} = 0 = D_{S} W_{\rm flux}$. After achieving the supersymmetric stabilisation of the axio-dilaton and the CS moduli, one has $\langle W_{\rm flux} \rangle = W_0$. However, the no-scale structure protects the K\"ahler moduli $T_\alpha$ which remain flat, and as a second step, they can be stabilised via including other sub-leading contributions, e.g. those arising from the whole series of $\alpha^\prime$ and string-loop ($g_s$) corrections; e.g. see \cite{Cicoli:2021rub} and references therein. At two-derivative order, these effects can be captured through corrections in ${\cal K}$ and $W$,
\bea
\label{eq:KandW}
& & \hskip-1cm {\cal K} = K_{\rm cs} -\ln\left[-\,i\,(S-\bar{S})\right]-2\ln{\cal Y}, \\
& & \hskip-1cm W = W_0 + W_{\rm np}(S, T_\alpha), \nonumber
\eea
where $K_{\rm cs}$ depends on CS moduli, while ${\cal Y}$ generically encodes the volume/dilaton pieces such that ${\cal Y} = {\cal V}$ at tree-level. Here ${\cal V} = \frac{1}{3!} k_{\alpha\beta\gamma} t^\alpha t^\beta t^\gamma$ where $k_{\alpha\beta\gamma}$ is the triple intersection number of the CY$_3$ and the 2-cycle volumes $t^\alpha$'s are related to the divisor volume $\tau_\alpha$ via $\tau_\alpha = \partial_{t^\alpha} {\cal V}$.

\noindent
{\bf Standard LVS:} The LVS scheme of moduli stabilisation considers a combination of perturbative $(\alpha^\prime)^3$ corrections to ${\cal K}$ \cite{Becker:2002nn} and a non-perturbative contribution to $W$ \cite{Witten:1996bn}. The minimal LVS construction includes two K\"ahler moduli with a so-called Swiss-cheese like volume form: ${\cal V} = \lambda_b \tau_{b}^{3/2} - \lambda_s \tau_{s}^{3/2}$ where $\lambda_{b} = {\sqrt2}/({3\sqrt{k_{bbb}}})$ and $\lambda_s ={\sqrt2}/({3\sqrt{k_{sss}}})$. The K\"ahler potential ${\cal K}$ is given by Eq.~(\ref{eq:KandW}) with ${\cal Y} = {\cal V}+{\xi}/({2 g_s^{3/2}})$ such that $\xi=-{\chi({\rm CY}_3)\,\zeta(3)}/({16\pi^3})$, and $\chi({\rm CY}_3)$ being the CY Euler characteristic and $\zeta(3)\simeq 1.202$. We note in passing that moduli stabilisation in LVS requires $\xi>0$ or equivalently $\chi<0$, i.e. a constraint $h^{1,1}< h^{2,1}$ on the Hodge numbers counting the K\"ahler- and CS-moduli.

Furthermore, non-perturbative superpotential contributions \cite{Witten:1996bn} can be generated by using rigid divisors, such as shrinkable del-Pezzo (dP) 4-cycles, or by rigidifying non-rigid divisors using magnetic fluxes \cite{Bianchi:2011qh, Bianchi:2012pn, Louis:2012nb}. In fact, in the presence of a `diagonal' del-Pezzo (ddP) divisor, the so-called `small' $4$-cycle of the CY$_3$ gives: $W= W_0 + A_s\, e^{- i\, a_s\, T_s}$, where $T_s = c_s - i \tau_s$ for $c_s$ being the $C_4$ axion, and the holomorphic pre-factor $A_s$ is considered as a parameter after fixing CS-moduli. Subsequently, after minimising the axion $c_s$, the leading order pieces in the large volume expansion are collected as \cite{Balasubramanian:2005zx}:
\bea
\label{eq:VlvsSimpl}
& & \hskip-0.5cm V_{\rm LVS} \simeq \frac{\alpha_1}{{\cal V}^3} - \,\frac{\alpha_2\tau_s}{{\cal V}^2}\, e^{- a_s \tau_s} +\frac{ \alpha_3\,\sqrt{\tau_s}}{{\cal V}}\, e^{-2 a_s \tau_s},
\eea
where $\alpha_i$'s are positive constants defined as $\alpha_1 = \frac{3}{4} \kappa\hat\xi |W_0|^2$, $\alpha_2 = 4 \kappa a_s |W_0| |A_s|$, $\alpha_3 = 4 \kappa a_s^2 |A_s|^2 \sqrt{2 k_{sss}}$, while $\kappa = g_s e^{K_{\rm cs}}/(8 \pi)$. This potential (\ref{eq:VlvsSimpl}) results in exponentially large $\langle{\cal V}\rangle$ determined by
\bea
\label{eq:}
& & \hskip-0.75cm \langle {\cal V} \rangle \simeq \frac{\alpha_2 \sqrt{\langle \tau_s \rangle}}{2\,\alpha_3}\, e^{a_s \langle \tau_s \rangle}, \, \, \langle \tau_s \rangle \simeq \hat\xi^{2/3} \, \left(\frac{9\,k_{sss}}{8}\right)^{1/3}.
\eea
Here $k_{sss} = 9 - n$ is the degree of the dP$_n$ divisor such that dP$_0 = {\mathbb P}^2$. Thus the LVS models based on Swiss-cheese CY with $h^{1,1}({\rm CY}) \geq 3$ have leading order flat directions which can act as promising inflaton candidate with a sub-leading potential. However, the constraints from K\"ahler cone conditions get crucially important, especially in large field models like Fibre inflation.


\noindent
{\bf Perturbative LVS:} With the inclusion of the so-called {\it log-loop} effects \cite{Antoniadis:2018hqy,Antoniadis:2018ngr,Antoniadis:2019doc,Antoniadis:2019rkh,Antoniadis:2020ryh,Antoniadis:2020stf} along with the BBHL corrections \cite{Becker:2002nn}, the K\"ahler potential ${\cal K}$ is given by Eq.~(\ref{eq:KandW}) with ${\cal Y} = {\cal V} +  \frac{\hat\xi}{2} + \hat\eta\left(\ln{\cal V} -1\right)$, where $\hat\xi = \xi/g_s^{3/2}$, and $\hat\eta = -g_s^2\hat\xi\,\zeta[2]/\zeta[3]$. Now, considering the tree-level superpotential $W_0$ for CS+dilaton moduli stabilisation, one gets the following scalar potential,
\bea
\label{eq:pheno-potV2}
& & \hskip-0.5cm V_{\rm pert LVS} = \frac{3\, \kappa\, \hat\xi}{4\, {\cal V}^3}\, |W_0|^2 + \frac{3 \, \kappa\, \hat\eta\, (\ln{\cal V} - 2)}{2{\cal V}^3}\,|W_0|^2.
\eea
This subsequently results in an exponentially large VEV for the overall volume modulus determined as,
\bea
\label{eq:pert-LVS}
& & \hskip-0.5cm \langle {\cal V} \rangle \simeq e^{a/g_s^2 + b}, \, \, a = \frac{\zeta[3]}{2 \zeta[2]} \simeq 0.37, \, \, b = \frac73,
\eea
justifying the name-- ``perturbative LVS", which similar to the standard LVS, corresponds to an AdS minimum with large $\langle {\cal V} \rangle$. For $g_s = \{0.1, 0.2, 0.25\}$, Eq.~(\ref{eq:pert-LVS}) results in $\langle {\cal V} \rangle \simeq \{7.6 \cdot 10^{16}, 95595, 3567\}$ respectively!


\section{Fibre inflation and K\"ahler cone bounds on the inflaton}


\vskip-0.4cm
\noindent
One of the simplest extensions of the minimal two-field standard LVS model is to consider the models based on K3-fibred CY$_3$ with $h^{1,1} = 3$ and having a diagonal dP divisor \cite{Cicoli:2016xae,Cicoli:2018tcq,Altman:2021pyc,Shukla:2022dhz,Crino:2022zjk}. In a suitable divisor basis $\{D_b, D_f, D_s\}$ having the `big' divisor ($D_b$), the `small' divisor ($D_s$) and the K3 divisor ($D_f$), the typical volume-form is given as:
\bea
\label{eq:fibre-vol-form}
& & \hskip-1cm  {\cal V} = \frac{k_{bbf}}{2} t^f \, (t^b)^2 + \frac{k_{sss}}{6}(t^s)^3 = \lambda_f \tau_b \sqrt{\tau_f} - \lambda_s \tau_s^{3/2},
\eea
where $\lambda_{f} = {1}/({2\sqrt{k_{bbf}}})$ and $\lambda_s ={\sqrt2}/({3\sqrt{k_{sss}}})$. It turns out that ${\cal V}$ and $\tau_s$ are fixed by the minimal standard LVS scheme while $\tau_f$ remains flat and serves as the inflaton field in the presence of string-loop corrections.

With the appropriate string-loop effects to the K\"ahler potential, one can generate sub-leading corrections to LVS potential which depend on the third modulus $\tau_f$ and helps in driving inflation. The two typical corrections known as `KK' and `winding' types are \cite{Berg:2004sj,vonGersdorff:2005bf,Berg:2005ja,Berg:2007wt,Cicoli:2007xp,Gao:2022uop}
\bea
\label{eq:Vgs-KK-Winding}
& & V_{g_s}^{\rm KK} = \kappa \, g_s^2 \, \frac{|W_0|^2}{16\,{\cal V}^4} \sum_{\alpha,\beta} C_\alpha^{\rm KK} C_\beta^{\rm KK} \left(2\,t^\alpha t^\beta - 4\, {\cal V} \,k^{\alpha\beta}\right), \nonumber\\
& & V_{g_s}^{\rm W} = - \frac{\kappa\,|W_0|^2}{{\cal V}^3} \, \sum_{\alpha=1}^3 \frac{C_\alpha^W}{t^\alpha}\,,
\eea
where $C_\alpha^{\rm KK}$ and $C_\alpha^W$ are some model dependent coefficients which can generically depend on the CS moduli. For the minimal Fibre inflation model, one ends up with the following effective potential for K3-fibre modulus $\tau_f$,
\bea
\label{eq:Vfibre-minimal}
& & \hskip-1cm V(\tau_f) = V_{\rm up} + \frac{|W_0|^2}{{\cal V}^2} \left(\frac{B_1}{\tau_f^2} - \frac{B_2}{{\cal V}\, \sqrt{\tau_f}} + \frac{B_3 \, \tau_f}{{\cal V}^2} \right),
\eea
where $V_{\rm up}$ denotes some appropriate uplifting needed to obtain the dS minimum, and $B_i$'s are some CS moduli dependent parameters to be considered as constants for inflationary dynamics occurring at sub-leading order.


\noindent
{\bf Constraints from the K\"ahler cone :} In the minimal Fibre inflation model based on CY$_3$ with $h^{1,1} = 3$, two moduli, namely ${\cal V}$ and $\tau_s$, are fixed by the standard LVS while the third one corresponds to the inflaton field $\tau_f$. The crucial thing to note here is the fact that although $\tau_f$ remains flat after fixing the overall volume ${\cal V}$, there exists a bound on the field range it can traverse during inflationary dynamics. This arises because of the fact that the K\"ahler cone condition (KCC): $\int_{C_i} J > 0$ holds for all the curves $C_i$ in the Mori cone of the CY$_3$ which typically translates into the following set of constraints:
\bea
\label{eq:KC-gen}
& & \hskip-1cm {\rm KCC:} \qquad p_{\alpha\beta}\,t^\beta > 0, \quad {\rm for \, \, some} \quad p_{\alpha\beta} \in {\mathbb Z},
\eea
where summation for $\beta$ runs in $h^{1,1}$ while $\alpha$ corresponds to the number of KCC which is $h^{1,1}$ for simplicial cases but can be more for non-simplicial cases. Depending on the values of $p_{\alpha\beta}$ in a given concrete model, there will be constraints on $t^\alpha$, and so on the 4-cycle volumes. For example, taking the volume-form (\ref{eq:fibre-vol-form}) and using $\tau_b = k_{bbf} t^b t^f$, $\tau_f = \frac{k_{bbf}}{2} (t^b)^2$, $\tau_s = \frac{k_{sss}}{2} (t^s)^2$, a generic KCC of the form $p_s t^s + p_b t^b + p_f t^f > 0$ translates into,
\bea
\label{eq:KC-gen}
& & \hskip-1cm p_f \left({\cal V} + \lambda_s \tau_s^{3/2} \right) + 2\sqrt2\, \lambda_b \,p_b {\tau_f}^{3/2} > 3 \lambda_s\, p_s \sqrt{\tau_s} {\tau_f}.
\eea
Considering the generic possibilities that $p_b, p_f$ and $p_s$ can have either zero or non-zero values, results in 7 possibilities for the triplet $p_T^i = \{p_b, p_f, p_s\}$ where at least one of the three takes a non-zero value. It turns out that $p_T^1 = \{p_b,0,0\}$, $p_T^2 = \{0,p_f,0\}$ and $p_T^3 = \{0,0,-1\}$ give $\{\tau_b > 0, \tau_f > 0, \tau_s > 0,{\cal V} > 0\}$. In the presence of exceptional divisor, there always exists a `diagonal' basis in which one KCC becomes $t^s < 0$ or equivalently $\tau_s > 0$ corresponding to $p_T^3=\{0,0,-1\}$. In fact, the Swiss-cheese form appears due to this generic condition which picks up the negative sign of $t^s = \pm \sqrt{2 k_{sss} \tau_s}$. Further, $p_T^4 = \{p_b, 0, p_s\}$ gives a lower bound on $\tau_f$ such that $(p_s^2 k_{sss}\tau_s)/(p_b^2 k_{bbf}) < \tau_f$. On contrary to these, $p_T^5 = \{p_b, p_f, 0\}$ results in $\tau_f \lesssim {\cal V}^{2/3}$ and $p_T^6 = \{0, p_f, p_s\}$ implies that $\tau_f \lesssim {\cal V}/\sqrt{\tau_s}$. These cases naturally arise in explicit models \cite{Cicoli:2017axo, Cicoli:2018tcq}. Thus one has  an upper bound for flat modulus $\tau_f$ in term of $\langle {\cal V} \rangle$ despite the LVS potential (\ref{eq:VlvsSimpl}) being independent of $\tau_f$. For $h^{1,1}({\rm CY})=3$, KCC consists of 3 or more constraints of type (\ref{eq:KC-gen}), so the only possibility to avoid the upper bound on $\tau_f$ is to have a CY with KCC having 3 triplets from $p_T^i$ for $1\leq i \leq 4$. However, a detailed analysis of the CY geometries with $h^{1,1} = 3$ arising from the Kreuzer-Skarke database \cite{Kreuzer:2000xy} shows that it does not happen \cite{Cicoli:2018tcq} and one always have at least one of the $p_T^5, p_T^6$, and  $p_T^7$ which give the bound. 

For numerical demonstration of the problem, one can take the model proposed in \cite{Cicoli:2016xae} in which $k_{bbf} = 18, k_{sss} = 1$ corresponding to a dP$_8$ divisor, and we find that the three conditions from KCC in the given divisor basis are: $0 < 2 \langle \tau_s \rangle < \tau_f < (3 \langle{\cal V}\rangle + \sqrt{2} \langle\tau_s\rangle^{3/2})^{2/3}$. Using the values $\langle {\cal V} \rangle = 10^3, \langle \tau_s \rangle = 3, \langle \tau_f \rangle = 60$ as mentioned for the benchmark model in \cite{Cicoli:2016xae}, we find that KCC conditions result in $6 < \tau_f < 208.4$. Using canonical field $\varphi$ via $\tau_f = e^{2\varphi/\sqrt{3}}$ leads to $\langle\varphi\rangle = 3.55$, and $\varphi < 4.63$ corresponding to the bound $\tau_f < 208.4$, and hence $\Delta\varphi \simeq 1$ while $\Delta\varphi \simeq 6-8 M_p$ is typically needed for a successful Fibre inflation \cite{Cicoli:2008gp}. Regarding the challenge one faces in global model building, it is worth to mention that this setup was the one among just a few models found suitable for Fibre inflation after scanning a total of 305 CY geometries with $h^{1,1} = 3$, and even this faces a problem. 

In fact such an upper bound exists even beyond the minimal $h^{1,1} = 3$ models, e.g. for a K3-fibred CY with $h^{1,1} = 4$ with a dP and a toroidal-like volume-form \cite{Cicoli:2017axo, Cicoli:2018tcq}
\bea
& & \hskip-1cm  {\cal V} = k_{123} t^1 t^2 t^3 + \frac{k_{sss}}{6} (t^s)^3 = \frac{\sqrt{\tau_1 \tau_2 \tau_3}}{\sqrt{k_{123}}} - \lambda_s \tau_s^{3/2}.
\eea
The generic KCC condition $p_{\alpha\beta} t^\beta > 0$ takes the form
\bea
\label{eq:KC-gen2}
& & \hskip-0.5cm \frac{p_1 \tau_2 \tau_3}{k_{123}{\cal Z}} + {\cal Z}\left(\frac{p_2}{\tau_2}+\frac{p_3}{\tau_3}\right) > 3 \lambda_s\, p_s \sqrt{\tau_s},
\eea
where ${\cal Z} ={\cal V} + 3 \lambda_s \tau_s^{3/2}$. An explicit example of this type has the following four K\"ahler cone conditions \cite{Cicoli:2017axo}:
\bea
\label{eq:KC4}
& & \hskip-1cm -t^s > 0, \, \, t^1 + t^s > 0, \, \, t^2 + t^s > 0, \, \, t^3 + t^s > 0,
\eea
which correspond to four sets of values for $p_T^i=\{p_1, p_2,p_3, p_s\}$ given as $\{0,0,0,-1\}$, $\{1,0,0,1\}$, $\{0,1,0,1\}$, and $\{0,0,1,1\}$. These four sets correspond to $\tau_s > 0$, $\tau_2\tau_3 \gtrsim {\cal V} \sqrt{\tau_s}$, $\tau_2 \lesssim {\cal V}/\sqrt{\tau_s}$, and $\tau_3 \lesssim {\cal V}/\sqrt{\tau_s}$ respectively.
From (\ref{eq:KC4}) it is evident that one of the reasons for this bound is the presence of the exceptional rigid divisor which is needed to induce the non-perturbative superpotential contribution for the standard LVS.


\section{Fibre inflation in Perturbative LVS}


\vskip-0.4cm
\noindent
The {\it perturbative LVS} provides the nice features of standard LVS without having any non-perturbative effects, and hence the need of an exceptional divisor can be simply bypassed, and therefore the K\"ahler cone conditions would not put a strong bound on the inflaton field range. We will demonstrate this feature by considering a concrete K3-fibred CY orientifold with toroidal-like volume.


\noindent
{\bf Global model:} We consider a CY$_3$ with $h^{1,1} = 3$ corresponding to the polytope Id: 249 in the CY database of \cite{Altman:2014bfa} which has been studied earlier in \cite{Gao:2013pra, Leontaris:2022rzj, Bera:2024zsk}. It is described by the following toric data:
\begin{center}
\begin{tabular}{|c|ccccccc|}
\hline
Hyp &  $x_1$  & $x_2$  & $x_3$  & $x_4$  & $x_5$ & $x_6$  & $x_7$       \\
\hline
4 & 0  & 0 & 1 & 1 & 0 & 0  & 2   \\
4 & 0  & 1 & 0 & 0 & 1 & 0  & 2   \\
4 & 1  & 0 & 0 & 0 & 0 & 1  & 2   \\
\hline
& $K3$  & $K3$ & $K3$ &  $K3$ & $K3$ & $K3$  &  SD  \\
\hline
\end{tabular}
\end{center}
The Hodge numbers are $(h^{2,1}, h^{1,1}) = (115, 3)$, the Euler number is $\chi=-224$ while the Stanley-Reisner ideal is ${\rm SR} =  \{x_1 x_6, \, x_2 x_5, \, x_3 x_4 x_7 \}$. The analysis of the divisor topologies using {\it cohomCalg} \cite{Blumenhagen:2010pv, Blumenhagen:2011xn} shows that the first six toric divisors are K3 surfaces while the seventh one is described by Hodge numbers $\{h^{0,0} = 1, h^{1,0} = 0, h^{2,0} = 27, h^{1,1} = 184\}$. Considering the divisor basis $\{D_1, D_2, D_3\}$ and the K\"ahler form $J = t^1 D_1+t^2 D_2+t^3 D_3$, the second Chern class $c_2({\rm CY})$ is given as $c_2({\rm CY}) = 5 D_3^2+12 D_1 D_2 + 12 D_2 D_3+12 D_1 D_3$, while $k_{123} = 2$ is the only non-zero triple intersection leading to ${\cal V} = 2 \, t^1\, t^2\, t^3 = \frac{1}{\sqrt{2}}\,\sqrt{\tau_1 \, \tau_2\, \tau_3}$ where $\tau_1 = 2 t^2 t^3, \tau_2 = 2 t^1 t^3, \tau_3 = 2 t^1 t^2$ and ${\cal V} = t^1 \tau_1 = t^2 \tau_2  = t^3 \tau_3$ as in toroidal case. The K\"ahler cone conditions are:
\bea
\label{eq:KC-torus}
& & \hskip-1.5cm \text{KCC:} \qquad  t^1 > 0, \quad t^2 > 0, \quad t^3 > 0.
\eea
Note that unlike KCC in Eq.~(\ref{eq:KC4}) corresponding to a CY$_3$ with a dP divisor \cite{Cicoli:2018tcq}, the KCC in Eq.~(\ref{eq:KC-torus}) do not put any other bound except $\tau_\alpha > 0$ and ${\cal V} > 0$. Therefore, the $\tau_3$ modulus is not heavily constrained if two of them are already fixed by some moduli stabilisation scheme.

\noindent
{\bf Brane setting:} For a given holomorphic involution, one needs to introduce D3/D7-branes and fluxes in order to cancel all the D3/D7 tadpoles. In fact, one can nullify the D7-tadpoles via introducing stacks of $N_a$ D7-branes wrapped around suitable divisors (say $D_a$) and their images ($D_a^\prime$). However, the presence of D7/O7 also contributes to the D3-tadpoles, which receive contributions from  $H_3/F_3$ fluxes, D3-branes and O3-planes. The resulting D3/D7 tadpole cancellation conditions are given as \cite{Denef:2004cf,Denef:2004ze,Blumenhagen:2008zz}:
\bea
\label{eq:D3D7tadpole}
& & \hskip-0.5cm {\bf \rm D7:} \,\, \sum_a\, N_a \left([D_a] + [D_a^\prime] \right) = 8\, [{\rm O7}],\\
& & \hskip-0.5cm {\bf \rm D3:} \,\, N_3 = \frac{N_{\rm O3}}{4} + \frac{\chi({\rm O7})}{12} + \sum_a\, \frac{N_a \left(\chi(D_a) + \chi(D_a^\prime) \right) }{48},\nonumber
\eea
where $N_3 \equiv N_{\rm D3} + \frac{N_{\rm flux}}{2} + N_{\rm gauge}$ such that $N_{\rm D3}$ is the net number of D3-brane, $N_{\rm flux} = (2\pi)^{-4} (\alpha^\prime)^{-2}\int_X H_3 \wedge F_3$ is the contribution from background fluxes and $N_{\rm gauge} = -\sum_a (8 \pi)^{-2} \int_{D_a}\, {\rm tr}\, {\cal F}_a^2$ is due to D7 worldvolume fluxes. Considering the involution $x_7 \to - x_7$ results in fixed point set $\{O7 = D_7\}$ without any $O3$-planes, and subsequently one can consider 3 stacks of $D7$-branes wrapping each of the three divisors $\{D_1, D_2, D_3\}$, and the D3/D7-brane tadpoles are nullified via $8\, [O_7] = 8 \left([D_1] + [D_1^\prime] \right) + 8 \left([D_2] + [D_2^\prime] \right)+ 8 \left([D_3] + [D_3^\prime] \right)$ resulting in $N_3 = 44$ ! This number appears to be large enough to accommodate the required flux choice leading to a suitable value of $W_0$ parameter. 

Now, in order to obtain a chiral visible sector on the D7-brane stacks wrapping $D_1$, $D_2$ and $D_3$, we need to consistently turn on worldvolume gauge fluxes of the form: ${\cal F}_i = \sum_{j=1}^{h^{1,1}} f_{ij}\hat{D}_j + \frac12 \hat{D}_i - \iota_{D_i}^*B$
with $f_{ij}\in \mathbb{Z}$. Here, the half-integer contribution is due to Freed-Witten anomaly cancellation \cite{Minasian:1997mm,Freed:1999vc}. However, given that the three stacks of D7-branes are wrapping a spin divisor K3 with $c_1({\rm K3}) = 0$, and given that the triple intersections on this CY are even, the pullback of the $B$-field on any divisor $D_\alpha$ does not generate a half-integer flux, and therefore one can appropriately adjust fluxes such that ${\cal F}_\alpha \in {\mathbb Z}$ for all $\alpha \in \{1, 2, 3\}$. We shall therefore consider a non-vanishing gauge flux ${\cal F}_3$ on the worldvolume of the $D_3$ divisor while considering ${\cal F}_1 = 0 = {\cal F}_2$. Subsequently, similar to \cite{Cicoli:2017axo}, the vanishing of FI parameter leads to
\bea
\label{eq:tau2=tau3}
& & \tau_1= q \, \tau_2, \quad {\rm where} \quad q = -q_{31}/q_{32}.
\eea


\noindent
{\bf Sub-leading corrections to scalar potential :} Given that there are no rigid divisors present, a priory this setup will not receive non-perturbative superpotential contributions \cite{Witten:1996bn} from instanton or gaugino condensation. The divisor intersection analysis shows that all the three $D7$-brane stacks intersect at ${\mathbb T}^2$ while each of those intersect the $O7$-plane on a curve ${\cal H}_9$ defined by $h^{0,0} = 1$ and $h^{1,0} = 9$. Further, there are no non-intersecting $D7/O7$ stacks and no $O3$-planes, and therefore this model does not induce the KK-type string-loop corrections to the K\"ahler potential \cite{Berg:2004sj,vonGersdorff:2005bf,Berg:2005ja,Berg:2007wt,Cicoli:2007xp,Gao:2022uop}. However, as the $O7/D7$ stacks intersect on non-shrinkable 2-cycles with size $t_\cap^i$, one will have string-loop effects of the winding-type \cite{Berg:2004sj,vonGersdorff:2005bf,Berg:2005ja,Berg:2007wt,Cicoli:2007xp,Gao:2022uop}: $V_{g_s}^{\rm W} = -\frac{\kappa |W_0|^2}{{\cal V}^3}\sum_i\frac{{\cal C}_{wi}}{t^i_\cap}$ where $C_{wi}$ are CS-moduli dependent coefficients. This CY has several properties similar to a toroidal case, however basis divisors being $K3$ implies that their second Chern numbers are non-zero, $\Pi(D_\alpha) = 24$,  unlike ${\mathbb T}^4$ (see \cite{Carta:2022web} for CY threefolds with a ${\mathbb T}^4$ divisor), leading to the higher derivative F$^4$ corrections to the scalar potential $V_{{\rm F}^4} \propto \Pi_\alpha t^\alpha$ \cite{Ciupke:2015msa}. Summarising all the contributions, we have \cite{Bera:2024zsk}:
\bea
\label{eq:Vfinal-simp}
& & \hskip-0.2cm V = V_{\rm up} +  \frac{{\cal C}_1}{{\cal V}^3} \left(\hat\xi - 4\,\hat\eta + 2\,\hat\eta \, \ln{\cal V}\right) \\
& & \hskip-0.25cm + \frac{{\cal C}_2}{{\cal V}^4} \biggl({\cal C}_{w_1}\tau_1 + {\cal C}_{w_2}\tau_2 + {\cal C}_{w_3}\tau_3 + \frac{{\cal C}_{w_4}\,\tau_1 \tau_2}{2(\tau_1 + \tau_2)} + \frac{{\cal C}_{w_5}\,\tau_2 \tau_3}{2(\tau_2 + \tau_3)} \nonumber\\
& &  \hskip-0.25cm + \frac{{\cal C}_{w_6}\,\tau_3 \tau_1}{2(\tau_3 + \tau_1)} \biggr) \, +  \frac{{\cal C}_3}{{\cal V}^3}\,\left(\frac{1}{\tau_1} + \frac{1}{\tau_2}+\frac{1}{\tau_3} \right) + \cdots,\nonumber
\eea
where ${\cal C}_1 = \frac{3}{4}\,\kappa\, |W_0|^2 = \frac{3}{4}{\cal C}_2,\, {\cal C}_3 = 24\, |\lambda|\,\kappa^2\, |W_0|^4/g_s^{3/2}$ such that $|\lambda| \simeq {\cal O}(10^{-4})-{\cal O}(10^{-2})$ \cite{Grimm:2017pid,Cicoli:2023njy}. The second piece of (\ref{eq:Vfinal-simp}) fixes ${\cal V}$ in perturbative LVS while one of the other two moduli are fixed via (\ref{eq:tau2=tau3}), leaving an effective single field potential for $\tau_3$ modulus which drives inflation.


\section{Inflationary dynamics} 
Using the slow-roll parameters $\epsilon_V \equiv V^{\prime2}/(2V^2)$ and $\eta_V \equiv V^{\prime\prime}/V$, the cosmological observables such as the scalar perturbation amplitude $P_s$, the spectral index $n_S$, and the tensor-to-scalar ratio $r$ are defined as
\bea
\label{eq:cosmo-observables}
& & \hskip-1cm P_s \simeq \frac{V_{\rm inf}^\ast}{24 \pi^2 \, \epsilon_H^\ast}, \, \, n_s \simeq 1 + 2 \, \eta_V^\ast - 6\, \epsilon_V^\ast, \, \, r \simeq 16 \epsilon_V^\ast,
\eea
where the observables are evaluated at the horizon exit $\phi = \phi^\ast$. The experimental bounds on these observables are : $P_s = 2.105\pm 0.030 \times 10^{-9}, n_s=0.9649\pm 0.0042$, and $r<0.036$ at a pivot scale $0.05\, {\rm M{pc}}^{-1}$
 \cite{Planck:2018jri}. The number of e-foldings $N_e$ is given as a sum three contributions \cite{Liddle:2003as,Cicoli:2017axo}:
\bea
\label{eq:cosmo-observables1}
& & \hskip-0.5cm N_e \simeq \int_{\phi_{\rm end}}^{\phi_\ast} \frac{V}{V^\prime} d\phi \simeq 57 + \frac{1}{4} \ln(r_\ast V_\ast) - \frac{1}{3}\ln\left(\frac{10V_{\rm end}}{m_{\inf}^{3/2}}\right),\nonumber
\eea
where $\phi_{\rm end}$ corresponds to end of inflation determined by $\epsilon_V = 1$ and $m_{\rm inf}$ is the inflaton mass. Also, it is worth noting that typically $N_e \simeq 50$ for Fibre inflation \cite{Cicoli:2017axo,Bhattacharya:2020gnk,Cicoli:2020bao}.

Using 
$\tau_1 = \tau_2$  for $q=1$ in (\ref{eq:tau2=tau3}) one is left with two K\"ahler moduli, and the overall volume ${\cal V}$ is fixed via perturbative LVS at leading order. Subsequently, using (\ref{eq:Vfinal-simp}) and considering $\tau_3 \equiv \tau_f = e^{2\varphi/\sqrt{3}}$ for the canonical field $\varphi$ and making the shift $\varphi = \langle \varphi \rangle + \phi$, one gets the following effective single field scalar potential for the shift $\phi$,
\bea
\label{eq:Vinf}
& & \hskip-0.9cm V = {\cal C}_0 \biggl({\cal C}_{\rm up} + {\cal R}_0 e^{-\gamma\phi} - e^{-\frac{\gamma}{2}\phi} +{\cal R}_1 e^{\frac{\gamma}{2}\phi} + {\cal R}_2 e^{\gamma\phi}\biggr),
\eea
where $\gamma=2/\sqrt{3}$ and the other parameters are:
\bea
\label{eq:Cis-new}
& & \hskip-0.5cm {\cal C}_0 =\frac{\sqrt{2}{\cal C}_2 \tilde{\cal C}_w}{\langle{\cal V}\rangle^3 e^{\frac{\gamma}{2}\langle\varphi \rangle}},  {\cal R}_0 = \frac{{\cal C}_3 e^{-\frac{\gamma}{2}\langle\varphi \rangle}}{\sqrt{2}{\cal C}_2\tilde{\cal C}_w}, \frac{{\cal R}_1}{{\cal R}_0} = \frac{\sqrt{2} e^{\sqrt{3}\langle\varphi\rangle}}{\langle{\cal V} \rangle}, \\
& & \hskip-0.5cm \frac{{\cal R}_2[\phi]}{{\cal R}_0} = \frac{{\cal C}_2 {\cal C}_{w3} e^{2\gamma\langle\varphi \rangle}}{{\cal C}_3\,\langle{\cal V} \rangle} \biggl[1+\hat{\cal C}_w \left(1+\frac{e^{\sqrt{3}(\phi+\langle\varphi\rangle)}}{\langle{\cal V}\rangle \sqrt{2}}\right)^{-1}\biggr],\nonumber
\eea
where $\tilde{\cal C}_w =-(4{\cal C}_{w1} + 4{\cal C}_{w2} + {\cal C}_{w4})/4$ and $\hat{\cal C}_w = ({\cal C}_{w5}+{\cal C}_{w6})/({2{\cal C}_{w3}})$. Further, we note that ${\cal C}_0, {\cal R}_0$ and ${\cal R}_1$ do not depend on the inflaton $\phi$ while ${\cal R}_2[\phi]$ have a sub-leading dependence. Subsequently, ${\cal C}_{\rm up} = 1-{\cal R}_0-{\cal R}_1-{\cal R}_2[0]$ is the uplifting required to achieve a dS minimum with small cosmological constant. The anti-D3 uplifting of \cite{Cicoli:2015ylx,Crino:2020qwk, AbdusSalam:2022krp} is not directly applicable to our case due to the absence of $O3$-planes, however other schemes, e.g. $T$-brane uplifting could be used \cite{Cicoli:2015ylx,Cicoli:2021dhg,Leontaris:2022rzj}. 

The first three terms of (\ref{eq:Vinf}) determine the minimum while the other two terms create the steepening. In fact, the first three terms correspond to Starobinsky type inflationary potential \cite{Starobinsky:1980te} (see  also~\cite{Brinkmann:2023eph}). 
Therefore we need to see if this inflation remains robust against the other sub-leading corrections and if there are any knock-on effects on inflation dynamics.
For this purpose, the main task is to engineer the parameters such that sufficient e-foldings are generated before the steepening starts through ${\cal R}_1$ and ${\cal R}_2$ pieces.

As seen from (\ref{eq:Cis-new}), the parameters ${\cal R}_1 \ll 1$ and ${\cal R}_2 \ll 1$ given that they are volume (${\cal V}$) suppressed, and this is to be exploited in finding a sufficiently long plateau. The slow-roll parameters do not depend on ${\cal C}_0$ as it is an overall factor seen from (\ref{eq:Vinf}), hence $n_s$, $r$ and $N_e$ can be determined purely by three parameters ${\cal R}_0, {\cal R}_1$ and ${\cal R}_2$. Although ${\cal R}_2$ depends on $\phi$, this dependence is suppressed by an extra volume factor, and can be made insignificant for the choice ${\cal C}_{w5} \simeq - {\cal C}_{w6}$, i.e. $\hat{\cal C}_w \simeq 0$. Along these lines, to begin with, ${\cal R}_2$ can be taken as a constant in order to understand the analytics of the leading order dynamics. Under this assumption, ${\cal R}_0 = (1+{\cal R}_1 + 2{\cal R}_2)/2 \simeq 1/2$ for setting $\langle\phi\rangle = 0$, i.e. when $\varphi$ modulus reaches its minimum. So one can study the inflationary dynamics from the potential (\ref{eq:Vinf}) and determine the range of ${\cal R}_1$ and ${\cal R}_2$ which could produce a long enough plateau to generate sufficient efoldings $N_e$ and suitable $(n_s, r)$ values, and then ${\cal C}_0$ is fixed by matching the scalar perturbation amplitude $P_s$. Following this strategy we take three steps for having an analytic understanding of the inflationary model:

\noindent{\bf Step-1:} To begin with we start by considering 
\bea
\label{eq:step-1}
& & {\cal R}_1 = 0 = {\cal R}_2,
\eea
and hence ${\cal R}_0 = 1/2$. Analyzing this case we find the following qualitative features of the model
\bea
& & \phi_{\rm end} \simeq 1, \quad \phi^\ast \simeq 6.5, \quad {\cal C}_0 \simeq 10^{-10},
\eea
where the order of the parameter ${\cal C}_0$ is determined by matching the scalar power spectrum amplitude $P_s$ at horizon exit $\phi^\ast$, while $\phi_{\rm end}$ corresponding to $\epsilon_V = 1$ is used to determine the total number of efoldings $N_e$ gained between the end of inflation and the horizon exit. A benchmark numerical model is given as below,
\bea
\label{eq:benchmark1}
& & \hskip-0.5cm {\cal C}_0 = 4.12 \cdot 10^{-10}, \quad {\cal R}_0 = \frac{1}{2}, \quad {\cal R}_1 = 0 = {\cal R}_2,\\
& & \hskip-0.5cm \phi_{\rm end} \simeq 1.03, \quad \phi^\ast \simeq 6.48, \quad N_e \simeq 55.7, \nonumber\\
& & \hskip-0.5cm P_s \simeq 2.1 \cdot 10^{-9}, \quad n_s \simeq 0.966, \quad r \simeq 0.006. \nonumber
\eea
These are the typical features of the Starobinsky-like inflationary potential \cite{Starobinsky:1980te,Brinkmann:2023eph}.

\begin{figure}[H]
\centering
\includegraphics[width=8.5cm]{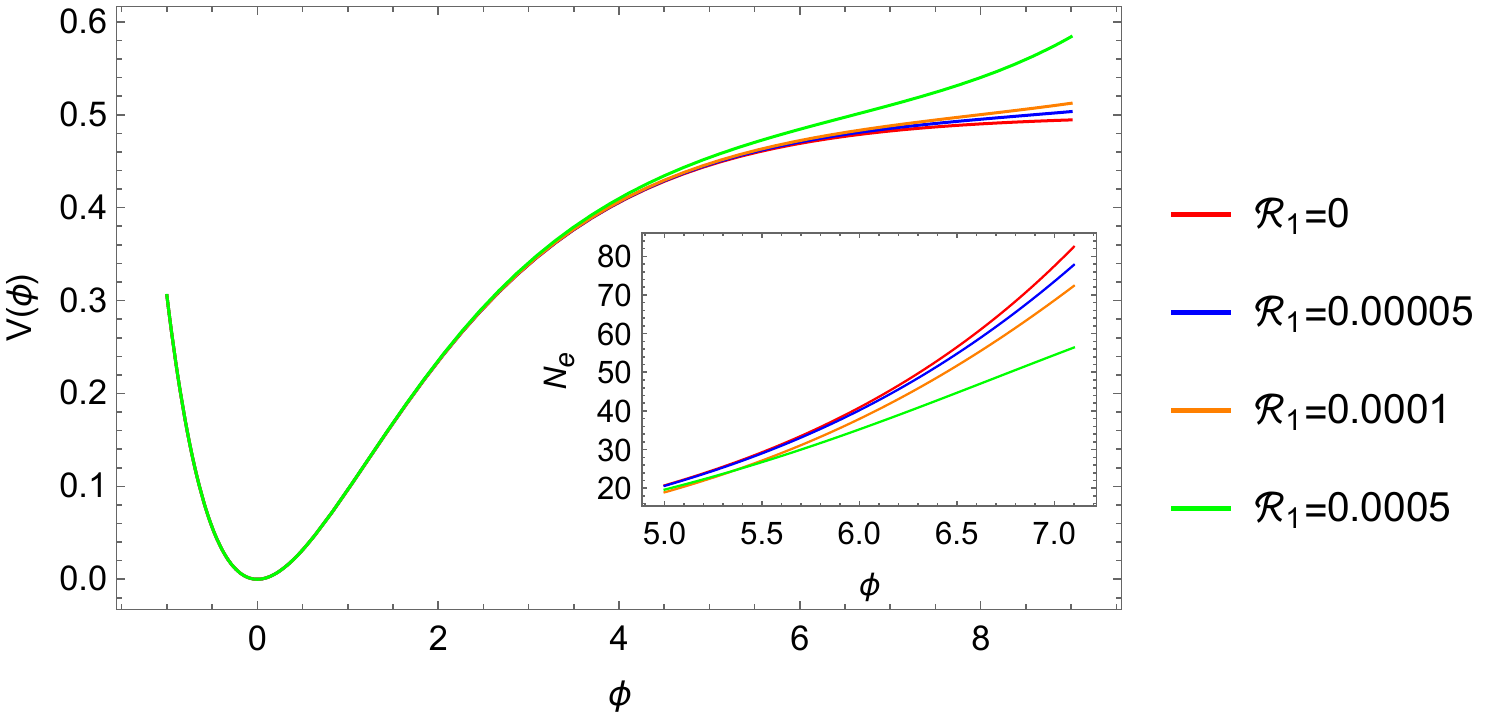}
\caption{$V(\phi)$ and $N_e$ are plotted for a set of ${\cal R}_1$ values}
\label{fig1}
\end{figure}

\begin{figure}[H]
\centering
\includegraphics[width=8.5cm]{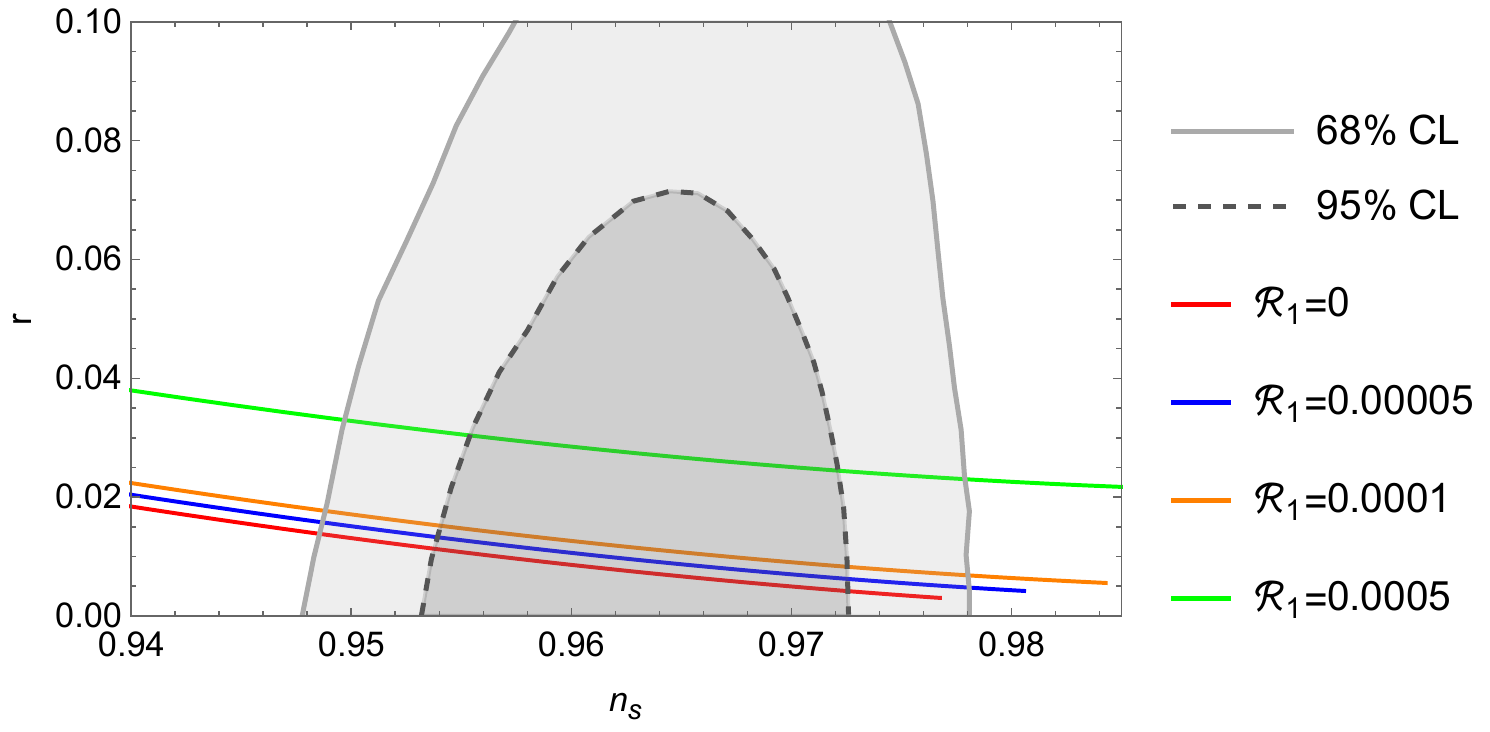}
\caption{$(n_s-r)$ plotted for a set of ${\cal R}_1$ values}
\label{fig2}
\end{figure}

\noindent{\bf Step-2:} Now as a second step we turn on the non-zero values for the ${\cal R}_1$ parameter while still keeping ${\cal R}_2 = 0$. This induces a steepening term in the inflationary potential which we need to control such that the earlier inflation is not spoiled. For this scenario we find that the model remains unaffected as long as $0 < {\cal R}_1 < 10^{-5}$. Increasing the ${\cal R}_1$ values starts developing the steepening piece but one continues to have a viable inflationary model with $0 \leq {\cal R}_1 < 0.0001$. However, we find that the larger ${\cal R}_1$ values destroy the plateau before sufficient e-folds are gained while maintaining to produce the cosmological observables within the required range. One set of numerical parameters presenting the features of this model is given as
\bea
\label{eq:benchmark2}
& & \hskip-0.5cm {\cal C}_0 = 6.07 \cdot 10^{-10}, \quad {\cal R}_1 = 10^{-4}, \quad {\cal R}_2 = 0,\\
& & \hskip-0.5cm \phi_{\rm end} \simeq 1.03, \quad \phi^\ast \simeq 6.39, \quad N_e \simeq 49.9, \nonumber\\
& & \hskip-0.5cm P_s \simeq 2.1 \cdot 10^{-9}, \quad n_s \simeq 0.969, \quad r \simeq 0.009. \nonumber
\eea
We can observe that despite pushing the $n_s$ values to the experimentally allowed upper limit, i.e. $n_s = 0.969$, the choice of parameter ${\cal R}_1 = 10^{-4}$ reduces the number of e-folding slightly below 50, and for ${\cal R}_1 = 5 \cdot 10^{-4}$ we have $N_e \simeq 34$ ! Therefore, it is better to consider smaller values of ${\cal R}_1$ to be on a safer side regarding the robustness of the model. For that reason we consider the following set of benchmark parameters as an outcome of the analysis of {\bf Step-2},
\bea
\label{eq:benchmark3}
& & \hskip-0.5cm {\cal C}_0 = 5.44 \cdot 10^{-10}, \quad {\cal R}_1 = 5\cdot10^{-5}, \quad {\cal R}_2 = 0,\\
& & \hskip-0.5cm \phi_{\rm end} \simeq 1.03, \quad \phi^\ast \simeq 6.37, \quad N_e \simeq 50.5, \nonumber\\
& & \hskip-0.5cm P_s \simeq 2.1 \cdot 10^{-9}, \quad n_s \simeq 0.966, \quad r \simeq 0.008. \nonumber
\eea
These feature are collectively seen from Fig.~\ref{fig1} and Fig.~\ref{fig2}. 

\noindent{\bf Step-3:} In the final step we consider ${\cal R}_1 = 5\cdot10^{-5}$, and subsequently explore for a suitable bound for ${\cal R}_2$ in order to know where steepening wrecks the inflationary model. For this scenario we find that the model remains unaffected as long as $0 < {\cal R}_2 < 10^{-8}$. Increasing ${\cal R}_2$ further starts developing the steepening piece but one can still have a (marginally) working inflationary model for $0 \leq {\cal R}_2 < 5\cdot 10^{-7}$, and larger ${\cal R}_2$ values destroy the plateau before sufficient e-folds are gained to produce the desired cosmological observables. One set of benchmark numerical parameters presenting the important features of this model is given as below
\bea
\label{eq:benchmark4}
& & \hskip-0.5cm {\cal C}_0 = 5.5 \cdot 10^{-10}, \quad {\cal R}_1 = 5\cdot10^{-5}, \quad {\cal R}_2 = 10^{-7},\\
& & \hskip-0.5cm \phi_{\rm end} \simeq 1.03, \quad \phi^\ast \simeq 6.37, \quad N_e \simeq 50.6, \nonumber\\
& & \hskip-0.5cm P_s \simeq 2.1 \cdot 10^{-9}, \quad n_s \simeq 0.967, \quad r \simeq 0.0085. \nonumber
\eea
Using this three step analysis we have narrowed down the relevant region to look for a viable inflationary model.

\noindent
{\bf Stringy parameters:} So far we have studied the inflationary dynamics from the potential (\ref{eq:Vinf}) by simply considering the ${\cal C}_0, {\cal R}_0, {\cal R}_1$ and ${\cal R}_2$ as constant parameters, without getting into the corresponding stringy parameter details. Now we show that using the various stringy parameters, similar numerical models can be produced. For that purpose, using (\ref{eq:Cis-new}) the values of stringy parameters can be typically estimated by  the following relations
\bea
\label{eq:stringy-parameters-gen}
& & \hskip-0.5cm e^{\frac{1}{2}K_{\rm cs}}\,|W_0| = \frac{2^{2/3}\, {\cal C}_0^{1/4}\, \sqrt{\pi}\, {\cal R}_0^{1/12}\, {\cal R}_1^{1/6}\, {\langle{\cal V} \rangle}^{11/12}}{3^{1/4}\, g_s^{1/8}\, |\lambda|^{1/4}},\\
& & \hskip-0.5cm \tilde{\cal C}_w = \frac{2\sqrt{3} {\cal C}_0^{1/2} {\langle{\cal V} \rangle}^{3/2} \sqrt{|\lambda|}}{g_s^{3/4}\, \sqrt{{\cal R}_0}}, \quad {\cal C}_{w3} = \frac{2 {\cal R}_0{\cal R}_2}{{\cal R}_1} \tilde{\cal C}_{w} 
, \nonumber\\
& & \hskip-0.5cm \langle{\cal V} \rangle = \frac{\sqrt{2}\, {\cal R}_0\, e^{\sqrt{3}\langle\varphi\rangle}}{{\cal R}_1}, \nonumber
\eea
where we recall that ${\cal R}_0 = (1+{\cal R}_1 + 2{\cal R}_2)/2 \sim 1/2$ under the assumption that ${\cal R}_0, {\cal R}_1$ and ${\cal R}_2$ are constant parameters where, in addition, being volume suppressed one anticipates that ${\cal R}_1 \ll 1$ and ${\cal R}_2 \ll 1$ are natural choices. Also one of the relation in (\ref{eq:stringy-parameters-gen}) shows that for the choice of ${\cal R}_1$ and ${\cal R}_2$ values presented in (\ref{eq:benchmark4}) leads to ${\cal C}_{w3} \simeq 0.002 \,\tilde{\cal C}_w$, i.e. a set of hierarchical values of the parameters $\tilde{\cal C}_w$ and ${\cal C}_{w3}$ are needed. One may assume it to be feasible given that these parameters generically depend on the complex structure moduli and hence should be tunable to a couple of order without a problem! 

\begin{figure}[h]
\centering
\includegraphics[width=8cm]{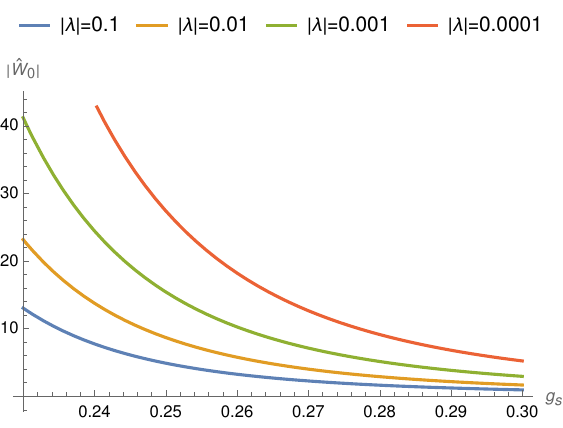}
\caption{$|\hat{W}_0|$ plotted for a set of $g_s$ values}
\label{fig3}
\end{figure}

\begin{figure}[h]
\centering
\includegraphics[width=8cm]{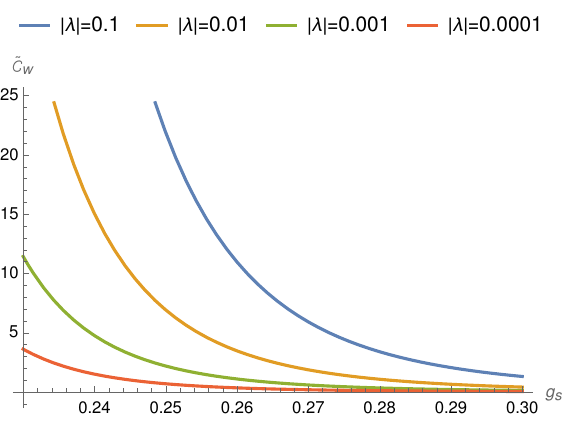}
\caption{$\tilde{\cal C}_w$ plotted for a set of $g_s$ values}
\label{fig4}
\end{figure}

\begin{figure}[h]
\centering
\includegraphics[width=8cm]{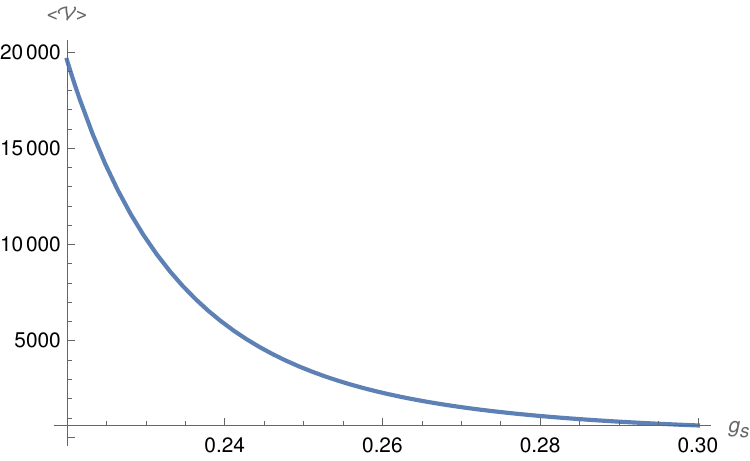}
\caption{$\langle {\cal V} \rangle$ plotted for a set of $g_s$ values in pert LVS}
\label{fig5}
\end{figure}

However we also note from (\ref{eq:stringy-parameters-gen}) that for the choice ${\cal R}_1 \simeq 10 {\cal R}_2$ one can reduce this hierarchy by an order, and subsequently can have $\tilde{\cal C}_w \simeq 10 {\cal C}_{w3}$ as ${\cal R}_0 \simeq1/2$. Having this in mind we present another benchmark model as below,
\bea
\label{eq:benchmark5}
& & \hskip-0.5cm {\cal C}_0 \simeq 5.49 \cdot 10^{-10}, \quad {\cal R}_1 \simeq 5\cdot10^{-6} \simeq 10\,{\cal R}_2,\\
& & \hskip-0.5cm \phi_{\rm end} \simeq 1.03, \quad \phi^\ast \simeq 6.35, \quad N_e \simeq 50.3, \nonumber\\
& & \hskip-0.5cm P_s \simeq 2.1 \cdot 10^{-9}, \quad n_s \simeq 0.967, \quad r \simeq 8.3\cdot10^{-3}. \nonumber
\eea
Now using the parameters of the working model presented in (\ref{eq:benchmark5}), the constraints in (\ref{eq:stringy-parameters-gen}) translate into the following ones,
\bea
\label{eq:stringy-parameters-num}
& & \hskip-0.5cm e^{\frac{1}{2}K_{\rm cs}}\,|W_0| \equiv |\hat{W}_0| \simeq \frac{0.001277\, {\langle{\cal V} \rangle}^{11/12}}{g_s^{1/8}\, |\lambda|^{1/4}},\\
& & \hskip-0.5cm \tilde{\cal C}_w \simeq \frac{0.000115 {\langle{\cal V} \rangle}^{3/2} \sqrt{|\lambda|}}{g_s^{3/4}}, \quad {\cal C}_{w3} \simeq \tilde{\cal C}_w/10, \nonumber\\
& & \hskip-0.5cm \langle{\cal V} \rangle \simeq 1.4\cdot 10^5\, e^{\sqrt{3}\langle\varphi\rangle} = 1.4\cdot 10^5\, {\langle\tau_f\rangle}^{3/2}, \nonumber
\eea
Note that we have defined $|\hat{W}_0|$ in order to keep track of the $K_{\rm cs}$ dependent piece. Also recall that the VEV of the overall volume $\langle {\cal V}\rangle$ and string coupling $g_s$ are related via (\ref{eq:pert-LVS}) in perturbative LVS. Therefore one can explore the favoured region of the parameter space $(|\hat{W}_0|, \tilde{\cal C}_w)$ for a given set of values of the string coupling $g_s$ and the higher derivative coupling $|\lambda|$. For this purpose $\hat{W}_0$, $\tilde{\cal C}_w$ as well as $\langle {\cal V} \rangle$ are plotted for a set of string coupling values as shown in Fig.~\ref{fig3}, Fig.~\ref{fig4} and Fig.~\ref{fig5} respectively. 

\noindent
From this analysis we make the following observations:
\begin{itemize}
\item 
For typical values of the string coupling $g_s \simeq 0.2-0.3$ and the higher derivative coupling $|\lambda| \simeq 0.0001-0.01$, demanding a large volume minimum such that $\langle {\cal V} \rangle \geq 10^4$ can lead to $|\hat{W}_0| \geq {\cal O}(10)$. Such values may be hard to consistently accommodate in a typical model where the D3-tadpole conditions result in smaller charges \cite{Denef:2004ze}, e.g. for our orientifold we have $N_3 = 44$ as discussed earlier. Therefore, in order to have $|\hat{W}_0| \simeq {\cal O}(1)$, the desired window for the overall volume should be $10^3< \langle{\cal V} \rangle < 10^4$ for the typical value of the $\lambda$ parameters, e.g. $0.001 < |\lambda| <0.01$. 

\item
We also observe that for a given $g_s$ value, the smaller values of $|\hat{W}_0|$ correspond to larger values of $|\lambda|$ parameter while it is reverse for the $\tilde{\cal C}_w$ parameter. Therefore the favoured region for having both of these parameters to be ${\cal O}(1)$ (and smaller than 10) corresponds to $g_s \simeq 0.24-0.26$ and $|\lambda| \simeq 0.001-0.01$. This range of string coupling typically leads to $\langle {\cal V} \rangle \simeq 10^3-10^4$.

\item 
However, the numerical estimates in (\ref{eq:stringy-parameters-num}) also show that the volume $\langle{\cal V}\rangle \simeq 10^3-10^4$  corresponds to negative values of $\langle \varphi\rangle$, i.e. fractional VEVs for $\langle \tau_f \rangle$ modulus. In this regard, we note that for trusting the overall effective supergravity analysis, the volume of each of the (four-cycle) modulus in string frame has to be larger
than the string scale $\ell_s$, and in particular one has to satisfy the following constraint for the 4-cycle volume moduli \cite{Cicoli:2017shd,Cicoli:2017axo,AbdusSalam:2020ywo}
\bea
\label{eq:fractional-taus}
& & \hskip-0.3cm \frac{\sqrt{\alpha^\prime}}{\tau_{\rm str}^{1/4}} \ll 1 \quad \Longrightarrow \quad (\tau_E \, g_s)^{1/4} \gg \frac{1}{2\pi} \simeq 0.16.
\eea
Here we note that we are working with the convention where $\ell_s = 2\pi \sqrt{\alpha^\prime}$ and the fact that the string-frame divisor volume and the Einstein-frame divisor volume are related via $\tau_E = \tau_{\rm str}/g_s$.

\end{itemize}

\noindent
Having learned the lessons from these numerical/graphical estimates we present a set of eight candidate models in Table \ref{tab_tab1}. Recalling that the current CY orientifold construction limits the net number of D3-brane charges to $N_3 = 44$ following from the D3 tadpole constraints and hence we keep $|\hat{W}_0| \lesssim 6$ \cite{Denef:2004cf,Denef:2004ze}. However a different CY orietifold may relax this condition significantly, e.g. those with large $N_3$ \cite{Crino:2022zjk,Shukla:2022dhz}. We find that the models M1-M6 which have $\langle {\cal V} \rangle \simeq 10^3-10^4$ can be correlated with the Fig.~\ref{fig3}, Fig.~\ref{fig4} and Fig.~\ref{fig5}. Also note that M6, which falls in the larger side of the volume, shows the need of larger $\tilde{\cal C}_w$. 

In addition we also present the models M7 and M8 corresponding to even a larger volume $\langle{\cal V}\rangle \simeq 1.5\cdot 10^4$ which might not be consistently possible to realize in the current global construction, however a model with $\langle{\cal V} \rangle = 15000$ and $g_s = 0.3$ has been realized in perturbative LVS with a de-Sitter minimum using T-brane uplifting \cite{Leontaris:2022rzj}. On these lines, the two models M7 and M8 give an extrapolated idea about the dependence of the various parameters, the moduli VEVs and the various cosmological observables. Also M7 and M8 show that having smaller values of $|\hat{W}_0|$ demands large $\tilde{\cal C}_w$ values and vice-versa which can be also observed by comparing the plots from the Fig.~\ref{fig3} and Fig.~\ref{fig4}. 

\begin{table}[h]
\begin{center}
\begin{tabular}{|c||c|c|c|c|c|c||c|c|} 
 \hline
Model & M1 & M2 & M3 & {\bf M4} & M5 & M6 & M7 & M8\\ 
\hline
\hline
$g_s$ & 0.28 & 0.27 & 0.26 & 0.25 & 0.24 & 0.23 & 0.224 & 0.224\\
$|\hat{W_0}|$ & 2 & 2 & 3 & 6 & 6 & 6 & 140 & 35 \\
$\tilde{\cal C}_w$ & 2 & 5 & 8 & 7 & 29 & 162 & 8 & 125 \\
${\cal C}_{w3}$ & 0.26 & 0.4 & 1.4 & 1 & 1 & 1 & 0.01 & 0.1\\
$\hat{\cal C}_{w}$ & 0 & 0 & 0 & 0 & 0 & 0 & 0 & 0 \\
$|\lambda|\cdot10^2$ & 3.5 & 8.5 & 5.9 & 1.3 & 5.2 & 28.3 & 0.015 & 3.79\\
\hline
$\langle{\cal V}\rangle$ & 1090 & 1549 & 2295 & 3567 & 5865 & 10305 & 14993 & 14993\\
$\langle\varphi\rangle$ & -3 & -3 & -3 & -3 & -3 & -3 & 0.1 & 0.1 \\
$\tau_{\rm str}^{1/4}$ & 0.31 & 0.30 & 0.30 & 0.30 & 0.29 & 0.29 & 0.71 & 0.71 \\
\hline
${\cal C}_0\cdot10^{10}$ & 5.5 & 4.6 & 4.93 & 4.42 & 3.95 & 3.90 & 5.54 & 5.41\\
${\cal R}_0$ & 0.5 & 0.5 & 0.5 & 0.5 & 0.5 & 0.5 & 0.5 & 0.5 \\
${\cal R}_1\cdot10^6$ & 3.6 & 2.5 & 1.7 & 1.1 & 0.67 & 0.38 & 56 & 56\\
${\cal R}_2\cdot10^7$ & 4.7 & 2.0 & 3.0 & 1.6 & 0.23 & 0.024 & 0.70 & 0.45\\
\hline
$\phi_{\rm end}$ & 1.03 & 1.03 & 1.03 & 1.03 & 1.03 & 1.03 & 1.03 & 1.03 \\
$\phi^\ast$ & 6.35 & 6.44 & 6.40 & 6.46 & 6.52 & 6.53 & 6.37 & 6.38\\
$N_e$ & 50.4 & 53.7 & 52.5 & 54.6 & 56.9 & 57.4 & 50.4 & 50.7\\
$P_s^\ast\cdot10^9$ & 2.13 & 2.10 & 2.12 & 2.09 & 2.08 & 2.10 & 2.08 & 2.11 \\
$n_s^\ast$ & 0.967 & 0.967 & 0.967 & 0.967 & 0.967 & 0.967 & 0.967 & 0.967 \\
$r^\ast\cdot10^3$ & 8.3 & 7.0 & 7.5 & 6.8 & 6.1 & 6.0 & 8.6 & 8.5\\
\hline
\end{tabular}
\end{center}
\caption{Stringy parameters and cosmological observables for eight candidate models}
\label{tab_tab1}
\end{table}

\noindent
Finally let us recall that we have analyzed the inflationary potential (\ref{eq:Vinf}) by considering the ${\cal C}_0, {\cal R}_0, {\cal R}_1$ and ${\cal R}_2$ as constant parameters. Although it is indeed true that ${\cal C}_0, {\cal R}_0$ and ${\cal R}_1$ do not depend on the inflaton $\phi$, the parameter ${\cal R}_2$ has a $\phi$ dependence with an additional ${\cal V}^{-1}$ suppression factor as seen from (\ref{eq:Cis-new}). This piece can be nullified via setting $\hat{\cal C}_w \simeq 0$ as considered earlier. However let us emphasize that this assumption $\hat{\cal C}_w \simeq 0$ is just a self-consistent simplification made in order to understand the analytics of the model in an easier way and it is not a necessary feature of the model itself, and therefore should not be considered as some additional tuning requirement. In fact a detailed numerical analysis shows that the earlier inflationary predictions remain true for natural $\hat{\cal C}_w$ values, e.g. a modified version of the model {\bf M4} described by the following parameters gives similar cosmological predictions:
\bea
\label{eq:final-model}
& & \hskip-1cm {\bf M4':} \quad |\hat{W_0}| = 6, \quad {\tilde{\cal C}_w} = 7, \quad {\cal C}_{w3} = 1/2, \quad \hat{\cal C}_w = 1, \\
& & \hskip-0.5cm |\lambda| = 1.27\cdot10^{-2}, \quad \langle\varphi\rangle = -3, \quad \tau_{\rm str}^{1/4} \simeq 0.3 > \frac{1}{2\pi}, \nonumber\\
& & \hskip-0.5cm g_s = 0.25, \quad \langle{\cal V}\rangle \simeq 3567,\nonumber
\eea
which result in ${\cal C}_0 \simeq 4.42 \cdot 10^{-10}, {\cal R}_0 \simeq 0.5, {\cal R}_1 = 1.1\cdot10^{-6}$ and 
\bea
{\cal R}_2 = \frac{0.142857 + 7.84174 \cdot 10^{-8} e^{\sqrt{3} \phi}}{910877 + e^{\sqrt{3} \phi}}.
\eea
Subsequently the cosmological predictions are:
\bea
& & \phi_{\rm end} \simeq 1.03, \quad \phi^\ast \simeq 6.47, \quad N_e \simeq 55.1, \\
& & P_s^\ast \simeq 2.13\cdot10^{-9}, \quad n_s^\ast \simeq 0.967, \quad r^\ast \simeq 6.7\times10^{-3}.\nonumber
\eea
For the model ${\bf M4'}$ given in (\ref{eq:final-model}), the effective supergravity approximations are justified by ensuring the mass hierarchies in Eq.~(\ref{eq:mass-hierarchy}) which translates into: 
\bea
\label{eq:mass-hierarchy-num}
& & \hskip-0.5cm m_{\rm inf} \sim 2.9\cdot10^{-6}, \quad H^\ast \simeq \sqrt{\frac{V_{inf}^\ast}{3}} \sim 8.4\cdot10^{-6}, \\
& & \hskip-0.5cm m_{3/2} = e^{\frac{1}{2}K}\, |W_0| \simeq \sqrt{\frac{g_s}{8 \pi}} \frac{|\hat{W_0}|}{{\cal V}} \sim 1.7\cdot10^{-4}, \nonumber\\
& & \hskip-0.5cm M_{\rm KK}\simeq \frac{\sqrt{\pi}}{\sqrt{\cal V} \, \tau_{\rm bulk}^{1/4}} \sim 3.8 \cdot10^{-3},\nonumber\\
& & \hskip-0.5cm M_s = \frac{\sqrt{\pi} \, g_s^{1/4}}{{\sqrt{\cal V}}} \sim 2.1\cdot 10^{-2}, \nonumber
\eea
where for $M_{\rm KK}$ for the bulk modulus $\tau_{\rm bulk} \sim {\cal V}^{2/3}$ \cite{Cicoli:2017axo}, and all the masses in (\ref{eq:mass-hierarchy-num}) are expressed in units of $M_p$. Here we note that $m_{3/2} < M_{\rm KK}$ results in a bound on $|\hat{W}_0|$ which is given as: $\sqrt{\frac{\kappa}{\pi}} \, |\hat{W}_0| < {\cal V}^{1/3}$ where $\kappa = g_s/(8\pi)$ \cite{Cicoli:2013swa,AbdusSalam:2020ywo}. Such a bound may be hard to satisfy for models with large $|\hat{W}_0|$ values, and having the smaller volumes! 

Before we conclude, let us also mention a couple of aspects which demand for further improvements in this model:

(i).~One may ask about the ``genuineness" of the {\it log-loop corrections} for the beyond toroidal cases as these corrections are motivated by the toroidal results \cite{Antoniadis:2018hqy,Antoniadis:2018ngr,Antoniadis:2019doc,Antoniadis:2019rkh,Antoniadis:2020ryh,Antoniadis:2020stf} and a direct computation for the CY orientifold case is missing. In fact the same is true for the usual KK and winding type string-loop effects which rely on a conjectured form motivated by the toroidal results \cite{Berg:2004sj,vonGersdorff:2005bf,Berg:2005ja,Berg:2007wt,Cicoli:2007xp}. In this regard it is worth emphasizing that some of these corrections have been independently (re-)derived in a field theoretic approach of \cite{Gao:2022uop} which can be taken as a good starting point towards understanding their genuineness for beyond toroidal models.

(ii).~In our approach we have considered the complex-structure moduli and the axio-dilaton to be stabilized by the leading order flux effects. We have also assumed the existence of a consistent uplifting mechanism to get a viable de-Sitter vacuum. In a concrete global construction, any attempt to address these aspects in a holistic way may face several subsequent issues, e.g. the required tuning in the various complex-structure moduli dependent coefficients, tadpole constraints etc. (e.g. see \cite{Crino:2020qwk,AbdusSalam:2022krp,Junghans:2022exo,Gao:2022fdi,ValeixoBento:2023nbv}), however these issues can be fully settled only after the explicit computations are known.



To conclude, we have presented a concrete global embedding for one of the most attractive inflationary models, namely Fibre inflation, realised in standard LVS, where it faces a generic challenge due to a geometric K\"ahler cone bound on the inflaton range. By demonstrating through concrete examples, we have argued that one of the main reasons for this bound is the presence of an exceptional del-Pezzo divisor which is an integral part of the standard LVS. As perturbative LVS does not require any non-perturbative superpotential contribution, such a requirement can be bypassed, and a viable inflation can occur via a combination of string-loop corrections and the higher derivative F$^4$-effects. This proposal consolidates the robustness of Fibre inflation beyond the standard LVS. However, there are some limitations and open challenges, especially regarding the dynamics of the complex structure moduli in determining the VEVs of the so-called model dependent parameters in the effective single field analysis, which remains to be addressed.


\noindent
{\underline{Acknowledgements}:} We thank Xin Gao for useful discussions. PS would like to thank the {\it Department of Science and Technology (DST), India} for the kind support.


\bibliographystyle{JHEP}
\bibliography{reference}


\end{document}
